\documentclass[fleqn,twoside]{article}
\usepackage{amsmath,amssymb, amsfonts}
\usepackage[headings]{espcrc2}

\def\braket#1{\mathinner{\langle{#1}\rangle}}
\DeclareMathOperator{\tr}{tr}

\newcommand{\sbraket}[1]{\lbrack #1\rbrack}
\newcommand{\dalpha}{\dot{\alpha}}
\newcommand{\dbeta}{\dot{\beta}}

\newcommand{\CP}{\mathbb{CP}}
\newcommand{\dbar}{\bar\partial}

\title{CSW rules for massive matter legs and glue loops}

\author{Rutger Boels\address{Niels Bohr Institute, Niels Bohr International Academy,\\ Blegdamsvej 17, DK-2100 Copenhagen, Denmark} and Christian Schwinn\address{Institut f\"ur Theoretische Physik E,\\ RWTH Aachen University, D - 52056 Aachen, Germany}\thanks{supported by the DFG Sonder\-forschungs\-bereich/Trans\-regio~9 ``Computergest\"utzte Theoretische Teilchenphysik''  and the BMBF grant 05HT6PAA} }

\begin{document}

\begin{abstract}
\noindent Cachazo-Svr\v{c}ek-Witten-type Feynman rules for massive matter scalar legs and pure glue loops are presented, obtained by deriving them directly from the space-time action. We comment on the derivation and some sample applications, in particular to calculating one loop effects in pure Yang-Mills theory. Furthermore, we derive CSW rules for effective Higgs-gluon couplings studied in the literature. In addition, it is shown how twistor techniques for deriving canonical field transformations explored for massless scalars extend to massless fermions.
\end{abstract}

\maketitle

\section{Introduction}
Witten's observations \cite{Witten:2003nn} about the twistor structure of scattering amplitudes have triggered many people to (re)consider a plethora of techniques for calculating scattering amplitudes, some of which may be relevant for actual experiments to be performed at the LHC. Apart from this direct physical motivation, there is also the appearance of unexpected simplicity within the results of the calculations. One of these is the by now old result of Parke and Taylor about the simplicity of the tree level amplitudes for gluons with two helicities unequal from the others \cite{Parke:1986gb}. This result is obscure when using textbook perturbation theory. Even better, one can calculate \emph{all} amplitudes in tree level Yang-Mills by combining MHV amplitudes as if they were vertices \cite{Cachazo:2004kj} using a simple prescription for off-shell momenta.  Usually simple results in complicated calculations in physics are the consequence of a symmetry.  Therefore we can try to elucidate the symmetry underlying the CSW rules by deriving them from the space-time Lagrangian. In a Lagrangian the symmetries of a problem should be central, which was historically the main reason why `actions' beat the 'analytic S-matrix' method. In recent years however many new technologies have pushed the latter program much beyond action based approaches in some areas. This note is not going to change that just yet, although there are already some interesting other applications of our formalism. 

In the work described here, CSW rules for Yang-Mills theory coupled mainly to a massive scalar are studied following \cite{Boels:2007pj,Boels:2008ef}, to which the reader is referred for details, definitions, conventions and a more complete bibliography. This study is motivated through supersymmetry in two ways. First of all, through a susy Ward identity \cite{Schwinn:2006ca} amplitudes with massive scalar legs are directly connected to amplitudes with massive quark legs. Second, through supersymmetric decomposition one can calculate the rational part of gluon amplitudes in pure Yang-Mills theory using massive scalars in the loop. Both of these are directly relevant for phenomenology, apart from the fact that the standard model contains an (uncolored) scalar. Furthermore, our derivations elucidate the exact connection between two seemingly different derivations of CSW rules in the literature and shows equivalence of the field transformations. This latter point is extended to massless fermions in this short note by studying the twistor lifting formulae for fermions. 

\section{The rules}
We study a scalar field $\phi$ in the fundamental representation of the gauge group described by the space-time Lagrangian
\begin{equation}
\label{eq:phi-lag}
\mathcal{L}=
\mathcal{L}(A)+
\mathcal{L}_\phi
 =\mathcal{L}(A)+ (D_\mu\phi)^\dagger D^\mu\phi-m^2\phi^\dagger \phi
\end{equation}
where $D_\mu=\partial_\mu-ig A$. By a field transformation method one can derive from this massive CSW rules in terms of the new gluon fields $B$ and $\bar{B}$ and scalar fields $\xi$ and $\bar{\xi}$,
\begin{multline}
 V_{\textrm{CSW}}(\bar B_1,B_2,\dots \bar B_{i},\dots, B_n)= \\
i2^{n/2-1}  \frac{ \braket{1i}^4}{\braket{12}\dots\braket{(n-1)n}\braket{n1}}  \label{eq:csw-glue}
\end{multline}
\begin{multline}
  V_{\textrm{CSW}}(\bar\xi_1,B_2,\dots \bar B_{i},\dots \xi_n)=   \\
-i2^{n/2-1}  \frac{ \braket{in}^2\braket{1i}^2}{\braket{12}\dots\braket{(n-1)n}\braket{n1}} \label{eq:csw-2phi} 
\end{multline}
\begin{multline}
  V_{\textrm{CSW}}( \bar\xi_1,B_2,\dots \xi_{i},\bar\xi_{i+1}\dots, \xi_n)=  \\
-i2^{n/2-2}  \frac{\braket{1i}^2\braket{(i+1)n}^2}{\braket{12}\dots\braket{n1}}
 \left(1+\frac{ \braket{1(i+1)}\braket{in}}{\braket{1i}\braket{(i+1)n}}\right) \label{eq:csw-4phi} 
\end{multline}
and an additional tower of vertices with a pair of scalars and an arbitrary number of positive helicity gluons that is generated from the transformation of the mass term:
 \begin{equation}
\label{eq:csw-mass}
V_{\text{CSW}}(\bar \xi_1,B_2,\dots, \xi_n)= i 2^{n/2-1}
\frac{- m^2 \braket{1n}}{
\braket{12}\dots\braket{(n-1)n }} 
\end{equation}
These vertices are combined into amplitudes by using the propagators
\begin{eqnarray}
D_{\bar\xi\xi}(p^2)= \frac{i}{p^2-m^2} &\quad& D_{\bar{B} B}(p^2)= \frac{i}{p^2}
\end{eqnarray}
Spinors corresponding to off-shell gluons and both on-shell and off-shell scalars are understood as usual in the CSW rules~\cite{Cachazo:2004kj} and are obtained from the momentum by contraction with an arbitrary but fixed anti-holomorphic spinor $\eta^{a}$:
\begin{equation}
\label{eq:csw-continue}
k_{\dot \alpha}=k_{\dot\alpha\alpha}\eta^{\alpha}.
\end{equation}
equivalently\footnote{up to a trivial field renormalisation}, they can be defined by
\begin{equation}\label{eq:spinormomentumtrick}
p^{\dot \alpha} p^{\alpha}=p^{\dot\alpha \alpha}
-\frac{p^2}{2p_+}\eta^{\dot\alpha}\eta^{\alpha}
\end{equation}
which will be used below. 

\subsection{Sample applications}
First of all, we have verified for a set of examples with small numbers of particles and in the general class of amplitudes with plus helicity gluons and two scalars that the above set of Feynman rules reproduce known field theory results. Furthermore, since these rules are holomorphic they elucidate the twistor structure of massive scalar amplitudes.  

\subsubsection*{BCFW recursion}
An interesting application of the above rules is to construct a simple proof of BCFW recursion relations for massive scalars which follows the lines of that for mass-less gauge fields using CSW rules \cite{Britto:2005fq}. This greatly simplifies the analysis compared to previous derivations. 

\subsubsection*{Glue loops}
As mentioned before, one can calculate pure glue one-loop amplitudes by tying a mass-less scalar coupled to Yang-Mills into a loop. However, by employing dimensional regularisation, the scalar does acquire an effective mass, which get integrated over. We verified that the above rules compute the correct four point all-plus one loop amplitude. Since collinear limits of positive helicity gluons are manifest in our rules, this fixes the form of the all-plus scattering amplitudes, up to the usual five point ambiguity.

The rules make explicit that some scattering amplitudes such as all-plus gluon ones are explicitly suppressed by a factor of $\epsilon$ in dimensional regularisation. The only way non-zero answer can arise then is if there is a UV pole in the calculation. Hence these amplitudes seem to arise as an anomaly of some sorts \cite{Bardeen:1995gk}. 

\subsubsection*{Ramifications for pure Yang-Mills loops}
A similar analysis directly for the pure glue Yang-Mills theory suggests a quantum completion of the CSW rules by adding a vertex of the form
\begin{multline}\label{eq:gluecomplet}
V_{\bar B_1,B_2,\dots\dots B_n}
= \\
\frac{-i  2^{n/2-1}\mu^2\braket{\eta 1}^3}{
\braket{12}\dots\braket{(n-1)n}\braket{n1}}
 \sum_{i=1}^{n-1}
\frac{\braket{(i+1)1}\braket{i(i+1)}}{\braket{\eta (1+i)}^2\braket{\eta i}}
\end{multline}
to the pure glue MHV vertex \eqref{eq:csw-glue}. This is obtained directly from a $\mu^2 \bar{A} A$ term in the lightcone Yang-Mills theory. This term has some desirable features, but it would be interesting to verify that only adding this term leads to all the correct scattering amplitudes. It does suggest however that even at higher loops there is an interesting stratification of contributions to the loop amplitudes which is in need of further clarification. One consequence would be that loop diagrams made only of CSW vertices always reproduce the leading poles in the $\epsilon$ expansion. 

\subsubsection*{Effective Higgs-gluon couplings}
The Higgs scalar is uncharged under the strong gauge group, but it does couple to fermions and fermions couple to glue. Hence, there is an effective coupling induced by loops. See e.g. \cite{Dixon:2004za} and references therein. In the approximation of heavy (top) quark mass, this interaction can be modelled by a local vertex formed by a dimension $5$ operator. This gives
\begin{equation}\label{eq:effhiggscoupl}
\tr  \frac{\alpha_s}{6 \pi v}\int dx^4 \tilde{H} F^2 
\equiv \tr \int dx^4 H F^2.
\end{equation}
with $v\sim 246 GeV$. In order to derive the CSW rules from this one should split $H = \phi+\phi^{\dagger}$ and write
\begin{equation}
\label{eq:higgs-action}
\int dx^4 H F^2 \sim \int dx^4 \phi F_+^2 + \phi^{\dagger} F_{-}^2 
\end{equation}
Using the twistor derivation as for pure glue Yang-Mills discussed below, it is easy to show that the CSW rules for pure Yang-Mills and the field $\phi$ consist of the vertices 
\begin{multline}\label{eq:higgsgluonstower}
    V_{n}(\phi_1,\dots\phi_{l-1},B_{l},\dots, \bar{B}_i,\dots, \bar{B}_{j},\dots, B_n)=\\
i 2^{n/2-1}  \frac{ \braket{ij}^4}{
  \braket{l(l+1)}\dots\braket{(n-1)n}\braket{nl}}
\end{multline}
and the CSW gluon propagator. Here $l$ runs from $0$ to $\infty$. The result for $l=1$ was obtained earlier \cite{Dixon:2004za} based on supersymmetric considerations and BCFW relations. In addition, one can add matter using the Noether procedure on twistor space and even more complicated couplings can be handled. 

Note however, for any scattering amplitude for the physical $H$ fields, separate amplitudes for $\phi$ and $\phi^{\dagger}$ need to be calculated. It would be nice to see if there is an application of the above observations to the exception of the scattering of a Higgs, pseudo-scalar Higgs ($\phi - \phi^{\dagger}$) and glue. 

\section{Deriving them rules}
Two completely differently motivated action based derivations of CSW rules have appeared in the literature. The basic idea in both is to consider a canonical field transformation which trivialises the self-dual part of the Yang-Mills theory. However, the details are very different. 
\subsubsection*{Canonical transformation}
In a development initiated by Mansfield \cite{Mansfield:2005yd} and pushed further by Ettle and Morris \cite{Ettle:2006bw}, the starting point is the lightcone Yang-Mills Lagrangian. In terms of helicities, the action roughly takes the form
\begin{equation}
L = L_{A \bar{A}} + L_{A  A \bar{A}} + L_{A \bar{A} \bar{A}} + L_{A A\bar{A} \bar{A}} 
\end{equation} 
after fixing lightcone gauge and integrating out the conjugate momentum to the lightcone gauge choice.  Mansfield's basic idea was to study a canonical field transformation which trivialises part of the theory,
\begin{equation}
L_{A \bar{A}} + L_{A  A \bar{A}} = L_{B \bar{B}}
\end{equation}
The theory on the left-hand side can be identified with self-dual Yang-Mills theory \cite{Chalmers:1996rq}. This equation and the canonicality constraint can be explicitly solved \cite{Ettle:2006bw} by expansion in terms of the new fields 
\begin{align}\label{eq:fieldexppureglue1}
A = \sum_i \Upsilon(p_1\ldots p_i) B_1 \ldots B_i \\
\bar{A} = \sum_{i,j} \tilde{\Upsilon}(p_1\ldots p_i) B_1 \ldots B_{j-1} \bar{B}_j B_{j+1} \ldots B_i \label{eq:fieldexppureglue2}
\end{align}
with relatively simple expressions for the transformation coefficients. The two remaining terms in the action are then in the new variables sums over vertices with the MHV field content. It has furthermore been verified up to $5$ gluons that the resulting expressions indeed sum to MHV vertices with precisely the off-shell continuation envisioned by CSW. A subtlety \cite{Ettle:2007qc} related to the LSZ reduction formula will be ignored in the following; this affects only the $\overline{\textrm{MHV}}$ 3 point tree level gluon amplitude in the current paper. 


A similar but more involved derivation can be followed for massless scalars \cite{Boels:2008ef} or for massive ones (treating the mass term as a vertex and using the same transformation as the massless case). Just as for pure glue, it is easy to see that vertices with MHV field content are generated. However, just as there it is generically hard to prove that they the vertices reduce to MHV expressions even for few-particle amplitudes. The exception to these are the transformation of $(\phi^{\dagger} \phi)^i$-type terms. Luckily, the other vertices can be obtained in a different way.  

\subsubsection*{Twistor actions}
In an action for four dimensional field theory one always integrates the Lagrangian over $\mathbb{R}^4$, since this is the space on which the Lorentz group acts linearly. It is known that it is very hard to extend this group (non-trivially) while retaining non-pathological scattering behaviour \cite{Coleman:1967ad}. The only known exceptions to this are supersymmetry and conformal symmetry. Focusing on the not-so-much studied latter option, one could ask the question if given a space-time action there is a Lagrangian on the space on which the conformal group in four dimensions acts linearly. This group, $SO(6) \sim SU(4)$ (temporarily ignoring signature), acts linearly on the complex space $\CP^3$, which is also known as \emph{twistor space}.  

To make this question more precise we will work in Euclidean space-time for technical reasons, as there one obtains schematically
\begin{equation}
\CP^3 = \mathbb{R}^4 \times \CP^1.
\end{equation}
Hence one useful way to view twistor space in Euclidean signature is as a so-called harmonic space: every field on it depends on coordinates $(x_{\alpha \dalpha},\pi_{\dbeta})$, where $\pi$ are homogeneous coordinates on the Riemann sphere $\CP^1$. These fields can be expanded in harmonics on the extra sphere (see appendix D of \cite{Boels:2008ef}). Inverting the expansion gives fields on space-time as integrals over the extra sphere coordinate of certain twistor fields. This is the basic observation which allows one to lift off-shell fields on space-time to twistor space. 

Since twistor space is strictly larger than space-time, there is a certain ambiguity in how to lift fields. This manifests itself in a twistor action in gauge symmetries which exist even for matter fields. Note that gauge symmetry on twistor space will naturally be larger than usual since the gauge parameter will depend on 6 instead of 4 real variables. The extra gauge symmetry can be fixed in several ways. There is for instance a natural 'space-time' gauge, which reduces the action to the space-time one. Fixing an axial gauge using a spinor $\eta^{\alpha}$ for all the fields however leads directly to the CSW rules \cite{Boels:2006ir}, including the full vertices. 

In CSW gauge the self-duality equations of Yang-Mills turn out to be simply linear. Moreover, as gauge transformations are simply \emph{linear} canonical transformations, the construction is in spirit already very close to the canonical transformation method. This can be made precise by deriving the canonical transformation coefficients directly from the twistor lifting formulae, as done in \cite{Boels:2008ef} for scalars and gluons.

\section{Fermions in the twistor approach}
Similar twistor techniques as worked out for scalars can also be applied to fermions\footnote{This section contains material not presented at Loops and Legs.} as in \cite{Boels:2007qn} where the action was obtained. However, there the field transformations for the fermions were not written explicitly. A canonical derivation of a field transformation for fermions which gives the action an apparent MHV field content has recently appeared in \cite{Ettle:2008ey} which follows the same lines as the one in \cite{Boels:2007pj,Boels:2008ef}. Since this is the harder part of comparing canonical and twistor approaches, it is straightforward to complete the comparison also for fermions by obtaining these results from the twistor action. In the appendix of \cite{Boels:2008ef}, it was shown in an example how to couple fields to Yang-Mills theory in the twistor action approach using the Noether procedure. The same procedure can be followed for fermions transforming in the fundamental and below we indicate briefly what changes for the lifting formulae compared to \cite{Boels:2008ef}, where all definitions may be found. 

The starting point is, again, 
\begin{equation}\label{eq:uncoupledfermion}
S_{\textrm{fermion}} = \int_{\CP^3}  d^4\!x dk \bar{\psi}_{\alpha}\left(\dbar^{\alpha} \psi^{0} + \dbar^0 \psi^{\alpha} \right) + \bar{\psi}^{0} \dbar_{\alpha} \psi^{\alpha} 
\end{equation}
for two as yet uncharged weight $-1$ and $-3$ twistor $(0,1)$ form fields $\psi, \bar{\psi}$ respectively and the natural volume form $dk$ on the sphere. This action is invariant under $\psi \rightarrow \psi + \dbar f^{-1}, \bar{\psi} \rightarrow \bar{\psi} + \dbar f^{-3}$. The gauge symmetry can be used to fix a gauge which reduces the above action to
\begin{equation}
S_{\textrm{fermion}} = \int d^4\!x \bar{\nu}_{\dalpha}(x) \partial^{\alpha \dalpha} \nu_{\alpha}(x)
\end{equation}
The exact relation between $\nu(x)$ and $\psi(x,\pi)$ fields follows from the twistor lifting formulae, derived below. In general the field $\psi_0$ is `pure gauge' by a cohomology argument on the Riemann sphere (or simply by inspecting the harmonic expansions of the field and the gauge symmetry), so unambiguously,
\begin{equation}
\psi_0(x,\pi_1) = \dbar^{(1)}_0 \int_{\CP^1} \frac{\psi_0(\pi_2)}{\braket{\pi_1 \pi_2}}
\end{equation}
One easy check is that under a gauge transformation $\psi \rightarrow \psi + \dbar f^{-1}$ left and right side transform identically. With this the field equation for $\bar{\psi}^{\alpha}$ reads,
\begin{equation}
\dbar_0 \left(\psi_\alpha(x,\pi_1) + \dbar_\alpha \int_{\CP^1} \frac{\psi_0}{\braket{\pi_1 \pi_2}} \right) =0
\end{equation}
and hence the combination in parenthesis is independent of the sphere coordinate $\pi_1$, 
\begin{equation}\label{eq:barnuformulauncoupled}
\psi_\alpha(x,\pi_1) + \dbar_\alpha \int_{\CP^1} \frac{\psi_0}{\braket{\pi_1 \pi_2}} = \nu_{\alpha}(x) 
\end{equation}
for some field $\nu_{\alpha}$. Plugging this back into the action immediately gives the other lifting formula,
\begin{equation}\label{eq:nuformulauncoupled}
\bar{\nu}_{\dalpha}(x) = \int_{\CP^1} \bar{\psi}_0 \pi_{\dalpha}
\end{equation}
This particular formula also easily follows from a harmonic expansion argument (or an educated guess!). 

Coupling the fermion action \eqref{eq:uncoupledfermion} to Yang-Mills gauge fields follows the lines of the appendix in \cite{Boels:2008ef} which leads to the results of \cite{Boels:2007qn}. Basically, the difference with the scalar action is the difference in weight on the twistor space of the fields. This exactly reproduces the (massless) supersymmetric Ward identities between the different matter field amplitudes if one calculates the MHV amplitude in CSW gauge. By a similar argument, coupling in the gauge fields changes \eqref{eq:barnuformulauncoupled} and \eqref{eq:nuformulauncoupled} into:
\begin{align} 
\nu_{\alpha}(x) &= H_1^{-1} \psi_\alpha(x,\pi_1) \nonumber \\ & + H_1^{-1}\left(\dbar_\alpha + B_{\alpha}\right) H_{1}\int_{\CP^1} \frac{H^{-1}_{2}\psi_0}{\braket{\pi_1 \pi_2}}  \label{eq:nutrafo} \\
\bar{\nu}_{\dalpha}(x) &= \int_{\CP^1} \pi_{\dalpha} \bar{\psi}_0 H 
\end{align}  
To derive the formulae obtained in the canonical approach to CSW diagrams, one can use the contraction with the projection on the physical states by $\frac{\eta^{\alpha}}{\sbraket{\eta p}}$ and $\frac{\eta^{\dalpha}}{\braket{\eta p}}$ respectively. Furthermore, in \eqref{eq:nutrafo} it is advantageous to evaluate the formula at $\pi=\eta$ for which by gauge choice $H(\eta)=1$. This yields
\begin{align}
\nu(x) &= i \frac{\eta^{\alpha}\eta^{\dalpha} p_{\alpha\dalpha}}{\sbraket{\eta p}} \int_{\CP^1} \frac{H_{2}\psi_0}{\braket{\eta \pi_2}}   \\
\bar{\nu}(x) &= \int_{\CP^1} \bar{\psi}_0 H \frac{\braket{\pi \eta}}{\braket{\eta p}}
\end{align}
Using the fact that in CSW gauge all fields along the fibres will occur with $\delta(\eta_{\alpha} \pi_{\dalpha} p^{\alpha \dalpha})$ functions on the fibres and the expansion of the frames $H$, these can be written as
\begin{align}
\nu(p) = &  - i \sum_{i=1}^{\infty} \sqrt{2}^{i-1} \frac{\braket{\eta p}}{\braket{\eta 1} \braket{1 2} \ldots \braket{(i-1) i}}  \nonumber \\
&  \left(B_1 \ldots B_{i-1} \psi (p_i)\right) \\
\bar{\nu}(p) = &   \sum_{i=1}^{\infty} \sqrt{2}^{i-1} \frac{\braket{\eta 1} }{\braket{1 2} \braket{2 3} \ldots \braket{i \eta}} \frac{\braket{1 \eta}}{\braket{\eta p}} \nonumber \\
&  \left( \bar{\psi} (p_1) B_2 \ldots B_{i}\right)
\end{align}
where we have used \eqref{eq:spinormomentumtrick} and the following identification of the normalisation of the fields,
\begin{eqnarray}
B_j \equiv - i \frac{B_{0,j}}{\sbraket{\eta j}^2 } & \bar{\psi}_j = -i \frac{\bar{\psi}_{0,j}}{\sbraket{\eta j}}  & \psi_j = - i\sbraket{\eta j} \psi_{0,j} \nonumber
\end{eqnarray}
Comparing this to \cite{Ettle:2008ey}, the above transformations are equivalent after translating their notation and normalisations into our conventions. In addition, the transformation of the field $\bar{A}$ in the lightcone approach gets additional contributions which can be derived just as for the scalars from the $\bar{B}_0$ field equation in the twistor equation, yielding exactly the extra terms in equation (2.67) in \cite{Ettle:2008ey} after a translation. 

Similar techniques can in principle be applied to obtain explicit field transformation coefficients for $\bar{A}$ (and all the other fields) from twistor actions for \emph{arbitrary} spin-$0$ and spin-$\frac{1}{2}$ matter content by similar arguments. Canonicality for instance is guaranteed by the fact a simple gauge transformation is studied, which would need lengthy confirmation in the current form of the canonical approach. Note that this class of theories includes the full standard model.

\section{Conclusions}
We have presented the first example of complete CSW-type rules for massive scalar matter. Although the results offer only a partial improvement over ordinary Feynman rules in terms of analytic computational efficiency, there are several interesting applications to other types of problem, such as the proof of on-shell recursions relations for massive particles and the structure of one loop amplitudes in pure Yang-Mills. Especially for the latter there are serious open and interesting questions remaining, in particular about the extend to which the vertex in equation \eqref{eq:gluecomplet} forms a complete quantum completion of pure Yang-Mills theory.  

For the canonical approach, it would be interesting to investigate whether the mass term can be incorporated into the field transformation. In addition, a first order version of the canonical approach starting from \cite{Chalmers:1996rq} would be interesting, as this is the starting point of the twistor approach. For the latter one natural question is if the perturbation series for the electroweak sector of the standard model can be simplified using the extra twistor symmetry. All the ingredients are there but still need a clever gauge choice to become effective. 

\section*{Acknowledgement}
The authors thank the organisers for the opportunity to present this material at the Loops and Legs conference.

\end{document}